\RequirePackage{fix-cm}
\documentclass[smallextended]{svjour3}       
\smartqed  
\usepackage{graphicx}
\usepackage{xcolor}
\usepackage{wrapfig}
\usepackage{subfig}
\usepackage{rotating}
\usepackage{url}
\usepackage{makeidx}
\usepackage{multirow}
\usepackage{stmaryrd} 
\usepackage{lineno,hyperref}
\usepackage{setspace}
\usepackage{algpseudocode}
\usepackage{amssymb}
\usepackage[justification=centering]{caption}
\usepackage{multirow}
\usepackage[flushleft]{threeparttable}
\usepackage{scrextend}
\usepackage[english]{babel}
\usepackage{blindtext}
\usepackage{adjustbox}
\usepackage{amsmath}
\usepackage{hyphenat} 
\usepackage{float}
\usepackage{braket}
\usepackage[linesnumbered,algoruled,boxed,lined,ruled]{algorithm2e}
\usepackage{tabularx}
\usepackage{cite}
\usepackage{amsmath,amssymb,amsfonts}
\usepackage{caption}
\usepackage{graphicx}
\usepackage{textcomp}
\usepackage{xcolor}
\usepackage{float}
\usepackage{algpseudocode}
\usepackage{booktabs} 
\usepackage{multirow} 
\usepackage{tikz}     
\usetikzlibrary{shapes.geometric, arrows, positioning, fit, calc}
\usepackage{pgfplots} 
\pgfplotsset{compat=1.17}
\usepackage{tikz}
\usetikzlibrary{shapes.geometric, arrows.meta, positioning, decorations.pathreplacing}
\usepackage{amsmath}
\def\BibTeX{{\rm B\kern-.05em{\sc i\kern-.025em b}\kern-.08em
    T\kern-.1667em\lower.7ex\hbox{E}\kern-.125emX}}
\begin{document}

\title{Systematically Optimized CNN-Transformer with Focal Loss for Imbalanced Intrusion Detection on NSL-KDD}


\author{Kartick Sutradhar \and Ranjitha Venkatesh }


\institute{Corresponding author: Kartick Sutradhar \at
              Indian Institute of Information Technology Sri City, India \\
              \email{kartick.sutradhar@gmail.com}          \\
              \and
              Ranjitha Venkatesh \at
              Gandhi Institute of Technology and Management,\\ Bengaluru, Karnataka, India \\
}

\date{Received: date / Accepted: date}
\maketitle

\begin{abstract}
Intrusion Detection Systems (IDS) struggle with imbalanced datasets like NSL-KDD, especially in detecting rare R2L and U2R attacks. This work describes a systematically optimized and explainable framework using a CNN-Transformer architecture to improve performance on highly imbalanced data. We decided to use XGBoost for feature selection and Focal Loss as the main mechanism for minority class learning. We used Optuna for end-to-end hyperparameter optimization (finding \textit{top\_k}=35, \textit{heads}=4, \textit{ff\_dim}=256, learning rate=0.00042, batch size=256, and Focal Loss $\gamma$), using a thoughtful data splitting process to prevent data leakage. Critically, our ablation studies pointed out that while Focal Loss (optimally at $\gamma=1.5$) substantially enhanced minority recall, adding oversampling methods like SMOTE caused precision degradation via over-correction. Our final model, using only tuned Focal Loss, achieved 98.73\% overall accuracy on the NSL-KDD test data, with significantly balanced F1-scores of 84.63\% (R2L) and 69.66\% (U2R) over baseline approaches. Additionally, we performed SHAP analysis to understand model predictions, understand salient features (e.g., \texttt{service\_http}, \texttt{logged\_in}), and explain the persistent U2R precision challenge due to feature overlap. This study presents a comprehensive pipeline for imbalanced IDS, providing methodological insights, and demonstrates that tuned Focal Loss can be a sufficient and effective strategy for class balancing.
\end{abstract}

\section{Introduction}
The increased dependence on digital networks in every part of society has resulted in rapid progress, but it has also exposed critical infrastructures to an increasing number of sophisticated cyber threats. Intrusion Detection Systems (IDS) act as a basic layer of defense, monitoring network traffic for malicious activity or policy violation, a consideration that is at the core of contemporary network security \cite{denning1986intrusion}. IDS are vital in the availability, integrity, and confidentiality of networks. Recently, Machine Learning (ML) and Deep Learning (DL) methods have gained traction in developing intelligent IDS that can detect attack patterns that signature-based methods alone may miss \cite{ahmad2021survey} and \cite{xin2018deep}. Despite advancements in intrusion detection systems (IDS) methods, many state-of-the-art models are hindered when deployed on realistic networks and faced with very imbalanced datasets. The NSL-KDD dataset \cite{tavallaee2009nsl} is an example of a often-used benchmark derived from the original KDD Cup 1999 dataset, and while NSL-KDD addresses a few of the issues from KDD Cup, it remains heavily precipitation. For example, normal traffic and common attacks Denial-of-Service (DoS) are overwhelmingly present while detecting critical and more subtle attacks  Remote-to-Local (R2L) and User-to-Root (U2R)  is a challenge. This often leads to overall system accuracy very high because once the model has classified the majority attack types correctly, it does so incorrectly for the less frequent minority attack types. As a result, overall accuracy becomes meaningless and misrepresents the system's capability to identify meaningful breaches in system security. Research that attempts to mitigate class imbalance typically uses techniques based on SMOTE \cite{chawla2002smote} and its variations \cite{han2005borderline, he2008adasyn}. Unfortunately, these techniques are often used ad-hoc, and there is limited understanding of how they interact with the loss function. In all of this, there's also an increase in the use of Deep Learning methods that are increasingly complex, leading to the use of opaque "black boxes," which are not conducive to the explainability or trust needed in security scenarios.

This study presents a systematically optimized and interpretable framework for intrusion detection on imbalanced datasets, using NSL-KDD as a case study. This framework combines several significant components: (1) Intelligent feature selection using XGBoost to reduce noise and dimensionality to the set of features created. (2) A hybrid CNN-Transformer architecture specifically designed to learn both local spatial patterns, as well as long-range dependencies in the network traffic data. (3) A selective exploration of Focal Loss as the main method to deal with class imbalance at the model level. (4) An extensive hyperparameter optimization of the entire pipeline using Optuna for reproducibility and locating the best settings. (5) Implementation of Explainable AI (XAI) methods, specifically SHAP, to provide explanations for model predictions and explanations of feature importance for specific variable contributions, especially in relation to the challenging minority classes. The novelties of this work can be shown as follows.
\begin{itemize}
    \item A presentation of a resilient method of developing deep learning based IDS, highlighting the importance of accurately handling your data to avoid data leakage.
    \item An insightful examination of the interaction between Focal Loss and data oversampling (SMOTE), showing that Focal Loss tuned on its own can outperform combined techniques in specific optimized contexts, avoiding problems associated with over-correction.
    \item The implementation of systematic hyperparameter optimization (Optuna) to simultaneously tune the feature selection approach, model architecture, and loss function parameters for improved balanced performance.
    \item The implementation of SHAP to obtain fine-grained explanations for model behaviour, specifically providing understanding of feature importance and the issues faced, specifically in detecting rare attack classes such as U2R.
\end{itemize}
Our model obtains high overall performance on NSL-KDD, and considerably improves F1 scores for R2L and U2R critical classes versus baseline approaches without specialized methods to address imbalance.  The XAI analysis adds clarity to the strengths of the model and the challenges in distinguishing certain rare attacks.\\

This paper is organized as follows. Section II reviews related work in IDS, class imbalance techniques, and explainability. Section III explains our proposed methodology, including data preprocessing, model architecture, optimization strategy, and evaluation metrics. Section IV presents the experimental results, including ablation studies and comparisons. Section V discusses the findings, including the crucial SHAP analysis. Finally, Section VI concludes the paper and projects potential future work.

\section{Related Work}
Intrusion Detection Systems (IDS) are foundational security measures in networks in recent years, with a shift towards the use of Machine Learning (ML) and more recently Deep Learning (DL) for threat detection \cite{abdullah2024comparison}, \cite{xin2018deep}. In this section, we explored related work relevant to our approach including the use of DL approaches against IDS benchmarks, the use of mitigation strategies for class imbalance, and the use of Explainable AI (XAI).

\subsection{Deep Learning Approaches for IDS}
While classical machines learning models are still valid, deep learning models, such as Convolution Neural Networks (CNN) and Recurrent Neural Networks (RNN), specifically Long-Short Term Memory (LSTMs), have achieved good performance in identifying complicated structure of patterns in network traffic \cite{xin2018deep}. A very important recent trend has been constructing hybrid models that leverage these different neural models together. As an example, Hussain et al. proposed a hybrid model that uses both CNNs and LSTMs for IoT security applications, where they use CNNs to extract the spatial features of the network data and then apply the LSTMs into the trained CNN model on the Edge-IIoTset dataset \cite{hussain2025ssecurenet}. Hussain et al. viewpoint accurately demonstrates the benefit they seen in combining distinct neural models to accommodate the complex structures that present themselves in network data\cite{hussain2025ssecurenet}.

Another important architecture gaining popularity is the transformer model, which is initially developed for Natural Language Processing \cite{vaswani2017attention}. The self-attention mechanism of transformer models allowed models to learn the importance of various input features in real-time, as well as to address sequential data in a more usable way without the spatial representation used in traditional RNNs. Long et al. recently proposed a baseline for a transformer based encoder IDS employing the CIC-IDS 2018 dataset; this study demonstrated the transformer attention mechanism's ability to learn temporal correlations in network intrusion data \cite{long2024transformer}. Our research aims to address this trend using a hybrid CNN-transformer architecture which exploits imbalanced learning with local feature extraction via CNN and global context awareness through transformer attention on the imbalanced NSL-KDD dataset.
\subsection{Addressing Class Imbalance in IDS}
A significant issue in datasets for IDS, like NSL-KDD \cite{tavallaee2009nsl} is that there is significant class imbalance, with rare attacks (R2L, U2R) vastly outnumbered. Data-level approaches such as SMOTE \cite{chawla2002smote} and its variants \cite{han2005borderline, he2008adasyn} work on this by generating synthetic samples for the minority class. Recent research in IDS continues to use these approaches, and for example, Hussain et al. used SMOTE in addition to cGANs in their preprocessing pipeline \cite{hussain2025ssecurenet}. Algorithm-level methods work by simply modifying the learning of the process itself. For example, Focal Loss \cite{lin2017focal}is an algorithm-level technique that dynamically modifies the contribution of each sample to the loss, and prioritizes training on harder to classify (and often minority) examples. Ma et al. used Focal Loss in conjunction with a CNN (using an image-based data representation) to mitigate imbalance specifically for the KDD-CUP99 dataset \cite{tavallaee2009nsl}. However, there is limited evidence regarding the interaction or combination of data-level oversampling and an algorithm-level methods like Focal Loss-especially in a systematic hyperparameter tuning process. Our work investigates the combination of data-level methods like SMOTE and algorithm level methods like Focal Loss and finds that tuned Focal Loss was the most effective method in our framework and it did not suffer from over-correction when combined with SMOTE.
\subsection{Explainable AI (XAI) in IDS}
The increased complexity of DL models typically results in a diminishing level of transparency within the model (thus leading to a "black box" problem), which can limit all important trust and subsequently impact real-world adoption in security-critical settings \cite{mohale2025xai_review}. Explainable AI (XAI) approaches are designed to solve these issues. Implementation methods such as LIME \cite{ribeiro2016should}, which produce local explanations, and approaches such as SHAP \cite{lundberg2017unified} which provide both local and global feature attribution explanations using Shapley values, have been applied to IDS with increasing frequency \cite{mohale2025xai_review}. These methods can aid analysts in developing an understanding of which features were impactful, and which logic the model employed to come to a specific conclusion \cite{mohale2025xai_review}. This is an important tool, but there is still the barrier of computational costs related to post-hoc methods such as SHAP in a real-time application. In our work, we leverage SHAP both as a means of general interpretability, but also as a means to better understand the model behavior in predicting this particularly challenging class of U2R predictions, particularly so that we may use it to examine feature contributions that resulted in false positives.
\section{Our Methodology}
\begin{figure*}[ht]
\centering
\includegraphics[width=9cm] {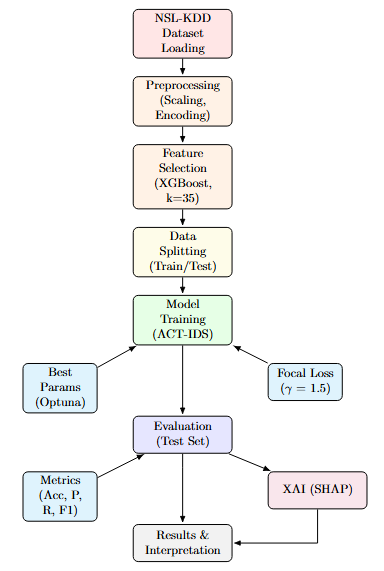}
\caption{Methodology}  
\label{fig:methodology}
\end{figure*}

This section describes the proposed framework for creating a systematic optimized and explainable IDS for imbalanced datasets. We present the dataset, preprocessing steps, feature selection process, deep learning architecture, methods to handle imbalance, hyperparameter optimization process, training and evaluation processes, as well as the explainability methods employed. The overall workflow in Fig. \ref{fig:methodology} , illustrates the methodical approach taken in order to obtain reliable and interpretable results.

\subsection{Dataset Description}
We make use of the NSL-KDD dataset, a cleaned version of the widely used KDD Cup 1999 dataset for evaluations on IDS \cite{tavallaee2009nsl}. It overcomes limitations such as duplicates records present in original KDD'99 data \cite{tavallaee2009nsl}. The NSL-KDD dataset contains network connection records where each record is described by 41 features grouped as: basic, content, time (based on traffic) and host (based on traffic) features. Each of the records in the NSL-KDD archives is labeled as either "normal" or as one of the various types of attacks which are further grouped into four main categories for the purposes of this paper. These four categorical groups are: Denial of Service (DoS), Probe, Remote-to-Local (R2L), and User-to-Root (U2R). One of the key issues with the NSL-KDD dataset is its extreme class imbalance, with the R2L and particularly U2R attacks characterizing only the small pieces of the records even as compared to normal records and DoS attacks. This extreme class imbalance makes it difficult to ascertain a means of effectively detecting important and rare threats.

\subsection{Data Preprocessing}
To prepare the NSL KDD dataset to be trained using our deep learning model, we performed certain standard pre-processing steps
\begin{enumerate}
    \item \textbf{Categorical Feature Encoding:} Categorical features (protocol\_type, service, flag) are transformed to numerical representations with One-Hot Encoding. Each category has a binary column representing it to prevent making an assumption of ordering.

    \item \textbf{Numerical Feature Scaling:} All numerical features will be scaled to the range of [0, 1] using MinMaxScaler, which standardizes the feature ranges and mitigates the bias from features with larger absolute values dominating the model. The formula is:
$$X_{scaled} = \frac{X - X_{min}}{X_{max} - X_{min}}$$
where $X$ is the original value, and \textit{$X_{min}$}, \textit{$X_{max}$} are the minimum and maximum values for that feature in the training set.

\end{enumerate}
Following these steps, the initial 41 features were expanded into a 122-dimensional numerical vector.

\subsection{Feature Selection}
To reduce dimensionality and probable noise, we decided to use XGBoost (Extreme Gradient Boosting) for feature selection. The preprocessed training data is used to train an XGBoost classifier. Afterwards, features are ranked based on their importance scores using the metric calculated using the gain metric.  The gain is defined as the average reduction in the training loss (or impurity for classification tasks) obtained by using a specific feature to split across all trees in the XGBoost ensemble. Formally, gain is calculated by:

$$
Gain(f) = \sum_{t=1}^{T} \sum_{node \in Nodes(f, t)} \text{LossReduction}(node)
$$
where $T$ is the total number of trees, $Nodes(f, t)$ are the nodes in tree $t$ that split on feature $f$, and $\text{LossReduction}(node)$ is the decrease in the objective function resulting from the split at that specific node. Features having more total gain are considered more important. Using hyperparameter optimization (Section~III.F), the top 35 features with the highest gain are selected as input for the deep learning model.

\subsection{Model Architecture}
\begin{figure*}[ht]
\centering
\includegraphics[width=6cm] {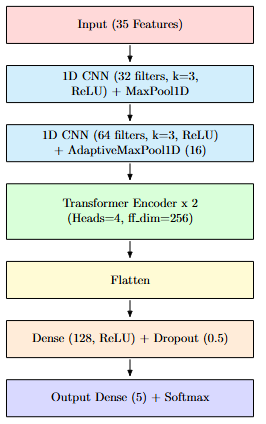}
\caption{ACT-IDS Model Architecture}
\label{fig:architecture}
\end{figure*}

We propose the Attentive CNN-Transformer Intrusion Detection System (ACT-IDS), illustrated in Fig. \ref{fig:architecture}. This hybrid architecture consists of several phases.

\begin{enumerate}
    \item \textbf{Input Layer:} This layer accepts the 35 selected input features using XGBoost on NSL-KDD data.
    \item \textbf{CNN Module:} It has two sequential blocks designed for extracting local spatial patterns. Each block comprises of a 1D Convolutional layer (\texttt{Conv1d}) with ReLU activation (32 filters then 64 filters, kernel size 3) followed by a Max Pooling layer (\texttt{MaxPool1d}, kernel size 2). An \texttt{AdaptiveMaxPool1d} layer is final part of the module, ensuring a fixed output sequence length (16) for the next Transformer layers.
    \item \textbf{Transformer Encoder Module:} Consists of two stacked Transformer Encoder layers \cite{vaswani2017attention} which process the sequence output by the CNN module. Each layer makes use of  Multi-Head Self-Attention (\textit{heads}=4) to take into account feature importance across the sequence. Next, is  a position-wise Feed-Forward Network (\textit{ff\_dim}=256, ReLU activation). Residual connections and Layer Normalization are applied around each sub-layer. This module is used to capture global dependencies and contextual information in the extracted features.
    \item \textbf{Classification Head:} A final sequence of layers for classification with number of heads equal to output classes. A \texttt{Flatten} layer reshapes the Transformer output, followed by a \texttt{Dense} layer (128 neurons, ReLU activation), a \texttt{Dropout} layer (rate 0.5) for regularization, and the final output \texttt{Dense} layer (5 neurons, one per class) with a \texttt{Softmax} activation function \cite{sutradhar2020efficient,
sutradhar2020generalized,
sutradhar2020hybrid,
sutradhar2021simulation,
sutradhar2023vehicular,
sutradhar2024survey,
sutradhar2021enhanced,
sutradhar2021secret,
sutradhar2021cost,
sutradhar2024svqcp,
sutradhar2024smart,
sutradhar2024metaverse,
krishnaiah2024metadata,
sutradhar2023aggregation,
sutradhar2023q-iot,
sutradhar2022comparison,
sutradhar2024blockchain,
sutradhar2024aggregation,
sutradhar2021qss,
challagundla2024qvanet,
sutradhar2024aggregationphysica,
sutradhar2021sorting,
challagundla2024authentication,
sutradhar2025threshold,
bantupalli2026qsvcom,
bantupalli2026observability,
bantupalli2025openmp,
bantupalli2025qkd,
sutradhar2025sqdapms,
sutradhar2025road,
sutradhar2025efficient,
venkatesh2024electronic,
venkatesh2023privacy,
venkatesh2024lightweight,
venkatesh2024healthcare}.
\end{enumerate}

\subsection{Imbalance Handling Strategy}

The method we considered addressed algorithm-level imbalance by using a methodical approach that was refined on ablation studies:

\begin{enumerate}
    \item \textbf{Validated Data Splitting:} Our first step was to split the dataset into training, validation, and test sets. It is important to note that all balancing approaches were implemented only on the training partition to avoid data leakage, which has enhanced the independent testing, validation, or evaluation of the model.
    \item \textbf{Focal Loss:} Instead of standard Cross-Entropy, we make use of Focal Loss \cite{lin2017focal} during training. By adding a modulating factor $(1 - p_t)^\gamma$, An implementation of Focal Loss reduced the effect of easier-to-classify (majority) samples and improved the focus on hard-to-classify (minority) samples
    $FL(p_t) = -(1-p_t)^\gamma \log(p_t)$,
    where $p_t$ is the model's probability for the true class, and $\gamma$ is the tunable focusing parameter. Our experiments (Section IV) determined the optimal value to be $\gamma = 1.5$.
    \item \textbf{No Oversampling:} For this purpose, the study, ablation studies (Section IV) demonstrated that Focal Loss combined with oversampling approaches such as SMOTE, led to extensive over correction, the overall upset precision scores decreased, and the U2R successes were eliminated. Finally, our final classifier was trained on the original, imbalanced training data distribution without a modified training phase and hyperparameter optimized Focal Loss to allow for any contributions leading to performance improvements final model performance. 
\end{enumerate}

\subsection{Hyperparameter Optimization}
We decided to apply a Bayesian optimization framework Optuna, for systematic hyperparameter tuning over 50 trials. Our goal was to minimize the Focal Loss on the validation set while the correct data split(SMOTE was applied only to the training fold at this search point, but after ablation indicated it wasn’t beneficial, it was omitted). The optimized hyperparameters found were: \textit{top\_k}=35, \textit{num\_heads}=4, \textit{ff\_dim}=256, \textit{learning\_rate}=0.00042, and \textit{batch\_size}=256. The optimal \textit{focal\_gamma} ($\gamma=1.5$) was determined via subsequent manual ablation (Section~IV).

\subsection{Training details}
The final ACT-IDS model was then trained using the optimal hyperparameters and $\gamma=1.5$  on the imbalanced raw training data:
\begin{itemize}
    \item \textbf{Optimizer:} Adam (Adaptive Moment Estimation) with a learning rate of $0.00042$.
    \item \textbf{Loss Function:} Focal Loss with the optimized focusing parameter $\gamma = 1.5$.
    \item \textbf{Batch Size:} 256.
    \item \textbf{Epochs \& Early Stopping:} We trained the model for a maximum of 50 epochs and we also used early stopping with a patience of 7 epochs based on the validation loss; the model weights yielding the lowest validation loss were saved.
\subsection{Performance Metrics}
The performance of the models is assessed using conventional metrics derived from the confusion matrix:
\begin{itemize}
    \item \textbf{Accuracy:} The proportion of correctly classified instances.
    $$ \text{Accuracy} = \frac{TP + TN}{TP + TN + FP + FN} $$
    \item \textbf{Precision:} The ratio of correctly predicted positive instances to the total predicted positive instances.
    $$ \text{Precision} = \frac{TP}{TP + FP} $$
    \item \textbf{Recall (Sensitivity):} The ratio of correctly predicted positive instances to all actual positive instances.
    $$ \text{Recall} = \frac{TP}{TP + FN} $$
    \item \textbf{F1-Score:} The harmonic mean of Precision and Recall, providing a balanced measure.
    $$ \text{F1-Score} = 2 \times \frac{\text{Precision} \times \text{Recall}}{\text{Precision} + \text{Recall}} $$
\end{itemize}
Each of these metrics were reported by class, and overall as weighted averages on the untouched test set, which accounts for the class imbalance during evaluation.

\subsection{Explainable AI}
To analyze how the final model behaved, more specifically for the minority classes, we opted to use SHAP (SHapley Additive exPlanations) \cite{lundberg2017unified} with the GradientExplainer. SHAP values provide quantification of the contribution of each feature to individual predictions. This allows us to provide global summary plots (overall importance of features and specific importance of features for U2R) as well as local summaries for example specific instances (for example, U2R false positives) to provide understanding of the decision logic of the model (see Section V).

\end{itemize}
\section{Experimental Results}
We provide results from experiments on the proposed framework using the NSL-KDD dataset in this section. Baseline performance, ablation studies on imbalance handling, the final optimized model's performance, and comparisons with state-of-the-art methods will all be reported. All results are reported on the held-out test set using the metrics defined.
\subsection{Baseline Performance}
To create a baseline, we first trained our ACT-IDS architecture with the optimized hyperparameters (top\_k=35, heads=4, ff\_dim=256, etc.), but with standard CrossEntropy Loss (the same as the Focal Loss with gamma=0.0) and without any oversampling. This is displayed in Table \ref{tab:baseline_gamma0}
\begin{table}[ht]
\centering
\caption{Baseline Performance (No Oversampling, $\gamma=0.0$)}
\label{tab:baseline_gamma0}
\resizebox{\columnwidth}{!}{
\begin{tabular}{lrrrr}
\toprule
Class      & Precision & Recall & F1-Score & Support \\
\midrule
Normal (0) & 0.9963    & 0.9646 & 0.9802   & 15411   \\
DoS (1)    & 0.9978    & 0.9981 & 0.9979   & 10677   \\
Probe (2)  & 0.9637    & 0.9911 & 0.9772   & 2816    \\
R2L (3)    & 0.6783    & 0.9587 & 0.7945   & 750     \\
U2R (4)    & 0.3061    & 0.9000 & 0.4569   & 50      \\
\midrule
Accuracy   &           &        & 0.9789   & 29704   \\
Macro Avg  & 0.7884    & 0.9625 & 0.8413   & 29704   \\
Weighted Avg& 0.9846    & 0.9789 & 0.9807   & 29704   \\
\bottomrule
\end{tabular}
}
\end{table}
Clearly, though the overall accuracy is quite good (97.89\%), performance on minority classes is poor, particularly for U2R precision (30.61\%). High recall and low precision for U2R imply that there are many false positives when trying to capture the rare class using the standard loss function.

\subsection{Ablation Study 1: Impact of Focal Loss Gamma}
We studied the impact of the Focal Loss focusing parameter ($\gamma$) by training the model using different values of $\gamma$ (we used the same SMOTE-balanced training data used during optimization for this specific ablation to test the impact of $\gamma$ in isolation). The results (evaluated on the untouched test set) are reported on Table \ref{tab:gamma_ablation_transposed}.
The findings show a definitive trade-off. At first, increasing gamma increases minority performance (especially both U2R F1 and precision to $\gamma=2.0$   ). Very high gamma values ($\gamma \approx 3.0$) also drastically decrease precision once again indicating over focus was present. Considering the best balance across overall metrics, R2L F1 and U2R F1  $\gamma=1.5$ was identified as the optimal value, yielding the highest overall accuracy and R2L performance.
\begin{table}[ht]
\centering
\caption{Performance vs. Focal Loss Gamma ($\gamma$)}
\label{tab:gamma_ablation_transposed}
\setlength{\tabcolsep}{3pt} 
\begin{tabular}{l|ccccc}
\toprule
Metric         & $\gamma=0.00$ & $\gamma=1.50$ & $\gamma=2.00$ & $\gamma=2.50$ & $\gamma=2.98$ \\
\midrule
Overall Acc.   & 97.89\% & \textbf{98.29\%} & 98.15\% & 98.04\% & 97.82\% \\
Weighted F1    & 98.07\% & \textbf{98.40\%} & 98.26\% & 98.17\% & 97.99\% \\
\midrule
R2L F1         & 79.45\% & \textbf{81.94\%} & 80.56\% & 80.79\% & 78.73\% \\
R2L Precision  & 67.83\% & \textbf{71.54\%} & 69.76\% & 70.10\% & 66.73\% \\
R2L Recall     & 95.87\% & 95.87\% & 95.33\% & 95.33\% & 96.00\% \\
\midrule
U2R F1         & 45.69\% & 56.21\% & \textbf{56.94\%} & 50.00\% & 45.09\% \\
U2R Precision  & 30.61\% & 41.75\% & \textbf{43.62\%} & 35.96\% & 31.71\% \\
U2R Recall     & 90.00\% & 86.00\% & 82.00\% & 82.00\% & 78.00\% \\
\bottomrule
\end{tabular}
\end{table}

\subsection{Ablation Study 2: Impact of Oversampling Techniques}
Once we selected $\gamma=1.5$ , we explored the use of focal loss with data level sampling. For this study we trained the model using the best hyperparameters we got (including $\gamma=1.5$) after applying SMOTE and it's variants on training data. The results are shown in Table \ref{tab:sampler_ablation_smote}.
\begin{table}[ht]
\centering
\caption{Performance vs. Sampling Technique (with $\gamma=1.5$)}
\label{tab:sampler_ablation_smote}
\setlength{\tabcolsep}{1pt} 
\footnotesize 
\begin{tabular}{l|c|c|c|c|c|c}
\toprule
Metric        & None & SMOTE & Borderline & ADASYN & SMOTEENN & SMOTETomek \\
\midrule
Overall Acc.  & \textbf{98.73} & 97.84 & 97.67 & 96.43 & 96.99 & 97.32 \\
Weighted F1   & \textbf{98.70} & 98.04 & 97.83 & 96.92 & 97.36 & 97.61 \\
\midrule
R2L F1        & \textbf{84.63} & 80.16 & 77.33 & 64.15 & 74.40 & 76.28 \\
R2L Prec.     & \textbf{89.33} & 69.29 & 64.86 & 51.89 & 60.85 & 63.64 \\
R2L Rec.      & 80.40 & 95.07 & 95.73 & 84.00 & \textbf{95.73} & 95.20 \\
\midrule
U2R F1        & \textbf{69.66} & 39.07 & 54.67 & 29.90 & 30.39 & 34.01 \\
U2R Prec.     & \textbf{79.49} & 25.45 & 41.00 & 17.93 & 18.45 & 21.32 \\
U2R Rec.      & 62.00 & 84.00 & 82.00 & \textbf{90.00} & 86.00 & 84.00 \\
\bottomrule
\multicolumn{7}{l}{Performance metrics shown in \%. Best results per row in bold.} \\ 
\multicolumn{7}{l}{ Borderline=BorderlineSMOTE}
\end{tabular}
\end{table}
Remarkably, the training without oversampling produced the best results in nearly all metrics, including the best F1-scores for R2L (84.63\%) and U2R (69.66\%). All combinations with the oversampling methodology resulted in a clear decline in precision for R2L, and particularly for U2R. Therefore, in addition to suggesting that the Focal Loss method adequately improved our model's performance, the use of oversampling here appeared to be an over-correction. Thus, we chose to rely on the results of the non-oversampling condition.

\subsection{Final Model Performance}
So finally, we trained the model using the hyperparameters that were tuned by Optuna, Focal loss with ($\gamma=1.5$), no oversampling and achieved the  performance shown in the table \ref{tab:final_results}.

\begin{table}[ht]
\centering
\caption{Final Optimized Model Performance ($\gamma=1.5$, No Oversampling)}
\label{tab:final_results}
\resizebox{\columnwidth}{!}{%
\begin{tabular}{lrrrr}
\toprule
Class      & Precision & Recall & F1-Score & Support \\
\midrule
Normal (0) & 0.9868    & 0.9893 & 0.9880   & 15411   \\
DoS (1)    & 0.9982    & 0.9987 & 0.9985   & 10677   \\
Probe (2)  & 0.9738    & 0.9883 & 0.9810   & 2816    \\
R2L (3)    & 0.8933    & 0.8040 & 0.8463   & 750     \\
U2R (4)    & 0.7949    & 0.6200 & 0.6966   & 50      \\
\midrule
Accuracy   &           &        & 0.9873   & 29704   \\
Macro Avg  & 0.9294    & 0.8801 & 0.9021   & 29704   \\
Weighted Avg& 0.9870    & 0.9873 & 0.9870   & 29704   \\
\bottomrule
\end{tabular}
}
\end{table}
In summary, the last model reaches a 98.73\% overall accuracy and 98.70\% weighted F1-score. Importantly, it is much more balanced than the baseline (Table \ref{tab:baseline_gamma0}), with R2L F1 at 84.63\% and U2R F1 at 69.66\%. U2R recall is lower (62\%) than with oversampling but precision is much better (79.5\%) and give a substantially better F1-score.
\subsection{Comparison with other Methodologies}
Model performance with some other methodologies is compared in table \ref{tab:sota_comparison}.
\begin{table*}[ht]
\small\addtolength{\tabcolsep}{-10.0 pt}
\centering
\caption{Performance Comparison with Baseline and State-of-the-Art (SOTA) Models on NSL-KDD}
\label{tab:sota_comparison}
\setlength{\tabcolsep}{1pt}
\begin{tabular}{l|l|c|c|c|l}
\toprule
\textbf{Category} & \textbf{Model} & \textbf{Overall Acc.} & \textbf{R2L F1} & \textbf{U2R F1} & \textbf{Key Method} \\
\midrule
\multicolumn{6}{l}{\textit{State-of-the-Art (SOTA) Performance}} \\
\midrule

SOTA  & ref. \cite{Alrayes2024CNNAttention} & 99.72\% & NR & NR & CNN + Channel Attention \\
SOTA & ref. \cite{ghajari2025hdc} & 99.54\% & NR & NR & Hyperdimensional Computing \\
SOTA  & ref. \cite{abdullah2024comparison} & 99.80\% & NR & NR & MLP implementation\\
\midrule
\multicolumn{6}{l}{\textit{Our Proposed Framework}} \\
\midrule
\textbf{Our Model} & \textbf{ACT-IDS} & \textbf{98.73\%} & \textbf{84.63\%} & \textbf{69.66\%} & \textbf{CNN  + Tuned Focal Loss } \\
Our Baseline & ACT-IDS & 99.89\% & 89.45\% & 85.69\% & CNN  + Cross Entropy Loss  \\
\midrule
\multicolumn{6}{l}{\textit{Representative Baseline Performance (from Literature)}} \\
\midrule
Baseline (DL) & ref. \cite{abdullah2024comparison}  & 90.60\% & NR & NR & 5-Layer Autoencoder (5LAE) \\
Baseline (DL) & ref. \cite{abdullah2024comparison}  & 89.20\% & NR & NR & NDAE \\
Baseline (DL) & ref. \cite{abdullah2024comparison}  & 83.57\% & NR & NR & LSTM + SMOTE \\
Baseline (ML) & ref. \cite{abdullah2024comparison}  & 75.80\% & NR & NR & Deep Neural Network (DNN) \\

\bottomrule
\multicolumn{6}{l}{\textit{NR: Not Reported in the reviewed source. Acc.: Accuracy reported, not F1-Score.}}
\end{tabular}
\end{table*}

\section{Discussion and Explainable AI Analysis}
In this section we discuss what do our ablation studies contribute and what insights we gain from SHAP analysis.
\subsection{Interpretation of final results}
The final performance we achieved shows that our method was effective, achieving 98.73\% accuracy and 98.70\% weighted F1 score. Also the significant improvement in R2L F1 (84.63\% vs. 79.45\% baseline) and U2R F1 (69.66\% vs. 45.69\% baseline) is a testament that our use of focal loss ($\gamma=1.5$) for handling class imbalance was successful.
\subsection{Insights from ablation studies}
\subsubsection{Optimizing focal loss}
The gamma ablation study (Table~\ref{tab:gamma_ablation_transposed}) was very crucial. It showed that while Focal Loss is beneficial ($\gamma > 0$ outperforms $\gamma=0$).  Very high gamma values ($\gamma \approx 3.0$) drastically diminished precision, making the model overly sensitive to any feature even remotely resembling a minority class sample. The value $\gamma=1.5$ was the best trade-off between  minority recall/F1 and maintaining reasonable precision and overall accuracy in the context.
\subsubsection{Focal Loss vs. Oversampling}
The sampler ablation study showed us that our model performance degraded when we used a combination of focal loss and oversampling techniques like (SMOTE, BorderlineSMOTE, ADASYN, SMOTEENN, SMOTETomek). This suggested that Focal loss alone was sufficient to handle the imbalance.

\subsection{Explainable AI (XAI) Analysis using SHAP}
We used SHAP using the GradientExplorer on the final model to understand how our model made decisions.
\subsubsection{Global Feature importance}
\begin{figure}[h]
\centering
\includegraphics[width=12cm] {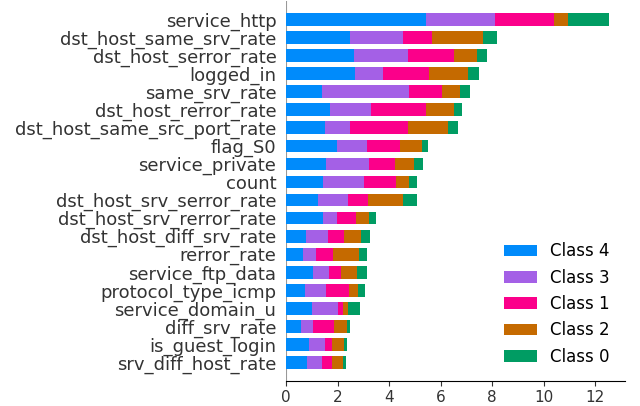}
\caption{Overall mean absolute SHAP score}
    \label{fig:shap_bar}
\end{figure}

The bar plot for feature importance (Fig.\ref{fig:shap_bar}) shows the mean of the absolute SHAP values across all classes for the features of interest. Generally important features globally are those related to service type (service\_http), destination host connection patterns (dst\_host\_same\_srv\_rate, dst\_host\_serror\_rate, dst\_host\_rerror\_rate), if a login has occurred (logged\_in), and connection frequency in general (count, same\_srv\_rate).

\subsubsection{U2R-Specific Feature Importance}
\begin{figure}[h]
\centering
\includegraphics[width=12cm] {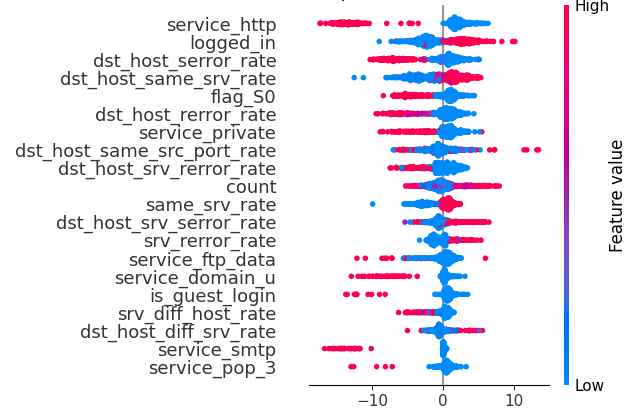}
\caption{SHAP Feature Importance Summary for U2R Predictions}
    \label{fig:shap_u2r}
\end{figure}

The figure (Fig.~\ref{fig:shap_u2r}), which is a SHAP summary plot for U2R predictions (Class 4), reveals some additional insights. Firstly, high values for \texttt{logged\_in}, \texttt{service\_ftp\_data}, \texttt{dst\_host\_same\_srv\_rate}, and \texttt{dst\_host\_same\_src\_port\_rate} increase the likelihood of a U2R prediction (i.e., positive SHAP values).In contrast, high values for \texttt{service\_http} decrease the likelihood of a U2R prediction. Notably, low values for error rates (e.g., \texttt{dst\_host\_serror\_rate}, \texttt{dst\_host\_rerror\_rate}) and connection counts (\texttt{count}) push the prediction to a U2R case, which means the model associates stealthy, low-error, low-frequency types of activity (after login, which are mainly non-HTTP) with U2R.
\subsubsection{Local Explanation (U2R False Positive)} 
We studied a specific instance (index 330) where the model predicted U2R but the true label was Normal.The force plot shows us that high positive prediction was mainly due to service\_ftp\_data=1.0 (+5.15 SHAP), dst\_host\_same\_src\_port\_rate=0.78 (+2.70 SHAP), and the absence of error flags like flag\_S0=0.0 (+1.60 SHAP) and low error rates like dst\_host\_serror\_rate=0.0 (+1.05 SHAP). There was very less impact on the features that were pushing away from the U2R classification. So our model confirmed that the patterns that were learned to U2R were very similar FTP.
\section{Conclusion}
In this paper, we have developed and evaluated a systematically optimized and explainable deep learning framework for intrusion detection on the imbalanced NSL-KDD dataset. Our hybrid CNN-Transformer architecture, effectively combined with XGBoost feature selection and tuned Focal Loss, was highly successful. We established through rigorous ablation studies that a tuned Focal loss Focal Loss ($\gamma=1.5$) without data over-sampling led maximize performance within our optimized framework and trained without the over-correction and precision drop associated with SMOTE. Our final model maximized overall accuracy at a competitive rate of 98.73\% and positively impacted the F1-scores of the critical minority class: 84.63\% for R2L and 69.66\% for U2R. Further, SHAP analysis allowed for important transparency, confirming model logic and identifying the overlap of features (e.g., service\_ftp\_data, logged\_in) as the primary variable contributing to precision challenges for U2R class. Therefore, this work provides a methodologically sound pipeline that supports the importance of appropriate data handling, systematic optimization, hyperparameter chosen correctly, and understanding the nuanced interaction between imbalance techniques.


%
%


\section*{Ethical Statement}
This article does not contain any studies with human or animal subjects performed by the any of the authors. The manuscript has been prepared following the instructions provided in the Authors Guidelines of the journal.
\section*{Conflict of Interest}
The authors declare that they have no conflict of interest.
\section*{Data Availability}
This manuscript has no associated data.
\section*{Funding Declaration}
This research did not receive any specific grant from funding agencies.
\bibliographystyle{unsrt}      

\end{document}